\documentclass[superscriptaddress, reprint, aps, prb]{revtex4-2}

\usepackage[T1]{fontenc}
\usepackage{graphicx, xcolor, nicefrac}
\usepackage{hyperref}
\usepackage{soul} 
\usepackage{amsmath}

\newcommand{\ea}{\emph{et al.}}
\newcommand{\pt}{phenyl-TOTA}
\newcommand{\mt}{methyl-TOTA}
\newcommand{\didv}{d$I/$d$V$}
\newcommand{\si}{Supplementary Information}

\newcommand{\ieap}{Institut f\"ur Experimentelle und Angewandte Physik, Christian-Albrechts-Universit\"at zu Kiel, 24098 Kiel, Germany}
\newcommand{\ioc}{Otto-Diels-Institut für Organische Chemie, Christian-Albrechts-Universit\"at zu Kiel, 24098 Kiel, Germany}
\newcommand{\csic}{Centro de Física de Materiales CFM/MPC (CSIC-UPV/EHU), 20018 Donostia-San Sebastián, Spain}

\begin{document}

\author{Behzad Mortezapour} \affiliation{\ieap} 
\author{Sebastian Hamer} \affiliation{\ioc} 
\author{Rainer Herges} \affiliation{\ioc} 
\author{Roberto Robles} \email{roberto.robles@csic.es} \affiliation{\csic}
\author{Richard Berndt} \email{berndt@physik.uni-kiel.de} \affiliation{\ieap}

\title{Orientational Order of Phenyl Rotors on Triangular Platforms on Ag and Au(111)}

\date{\today}

\begin{abstract}
We investigated trioxatriangulenium functionalized with phenyl (\pt) on the (111) surfaces of Ag and Au using low-temperature scanning tunneling microscopy (STM) and density functional theory (DFT).
On Ag(111), the molecules form hexagonal arrays, and on Au(111), honeycomb patterns are also observed.
The orientations of the phenyl moieties are resolved on both substrates.
On Ag(111), the orientations are parallel within a row and they differ by approximately $60^\circ$ between adjacent molecular rows, and STM images suggest dimerization of the molecules.
DFT calculations for Ag(111) reveal that van der Waals interactions dominate this system.
The optimized structure matches the experimental pattern, and the simulated STM images exhibit apparent dimerization.
This dimerization results from an asymmetry of the phenyl wavefunction, which reflects intramolecular hydrogen bonding between the ligand and an oxygen atom within the triangulenium platform. 
The orientation of the phenyl moieties is explained by the interaction of each phenyl moiety with its triangulenium platform combined with the direct long-range interaction between phenyl moieties across molecules.
\end{abstract}

\maketitle


The adsorption of molecules onto single-crystalline surfaces in vacuum or at liquid–solid interfaces often results in the self-assembly of decorative and potentially useful two-dimensional patterns.\cite{otero2011molecular, gutzler2011kinetics, goronzy2018supramolecular}
Understanding the patterns, which depend on various molecule-molecule and molecule-substrate interactions, as well as the detailed parameters of the preparation, is a formidable task.
Platforms in the shape of an equilateral triangle, deposited on a hexagonally symmetric substrate may seem to be a comparatively simple case. 
Indeed, various networks with honeycomb or hexagonal symmetries have often been observed.\cite{lackinger2005self, nath2006rational, ye2007unified, shen2014triangular, bao2017self, yang2020surface, fang2020stereospecific, feng2021boronic}
To add functionality to these arrays, molecules that serve as a platform for various ligands have been used.
In particular, the triangular compound trioxatriangulenium (TOTA) (inset of Fig.~\ref{intro} (a)) forms hexagonal or honeycomb arrays on the (111) surfaces of Ag and Au and can easily be functionalized with various axial ligands that stand vertically on the TOTA plane.\cite{kuhn2011adlayers, otte2014ordered, jasper2017conductance, jasper2018stability,jasper2019high, jasper2020rotation, jasper2020coverage}
The functional subunits in these arrays are separated by a distance of 1.01 to 1.04\,nm.
\footnote{When a bulky ligand with lateral dimensions exceeded those of the TOTA platform were used, other patterns have been observed.\cite{otte2014ordered}}

Phenyl-TOTA, which we investigate here, is particularly interesting because interactions between aromatic molecules are relevant in many areas of chemistry and engineering. \cite{Ninkovic2020}
The model case of benzene-benzene interaction has been computationally studied at various theoretical levels. \cite{Podeszwa2006, Lee2007, Bludsky2008, Ninkovic2020, Herman2023, Czernek2024}
While the detailed geometric arrangement of the molecules is known to play a role, the interaction energy at nm distance is expected to be of the order of few meV \cite{Czernek2024}.

In the present case, however, the functionalization with a phenyl subunit has a striking effect.
Using scanning tunneling microscopy (STM), the orientation of the phenyl moiety is resolved.
We find a striped phase of alternating rows that select two of three phenyl orientations differing by $\approx 60^\circ$.
In addition, the STM images suggest a dimer-like pairing of the molecules.
Density functional theory (DFT) calculations including van der Waals corrections, reproduce several important experimental findings and lead to the following interpretation:
the van der Waals interaction is by far the strongest interaction in this system.
In fact, it decreases the distance between the molecules and the substrate, and governs the orientation of the phenyls.
The calculated lowest energy structure matches the experimental pattern and leads to apparent dimerization in STM images.
This dimerization is due to the distortion of the phenyl wavefunction that results from the intramolecular hydrogen bonding between the ligand and an oxygen atom within the triangulenium platform.
The regular pattern of phenyl orientations results from the combination of two effects: the intramolecular hydrogen bonding, which determines three possible orientations of each ligand on the triangulenium platform, and the direct long-range interactions between the ligands, which determines their relative orientation.

\section*{Results and Discussion}

\subsection{Experimental Results on Ag(111)}

\subsubsection{Molecular Pattern on Ag(111)}

\begin{figure}[h!]
\includegraphics[width=1.0\linewidth]{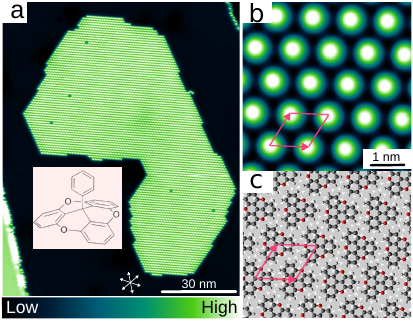}
\caption{
(a) Overview topograph ($V=0.15$\,V) of a submonolayer amount of \pt\ on Ag(111).
The inset shows a scheme of \pt\@. 
(b) More detailed topograph ($V=0.7$\,V) from the interior of a molecular island.
A unit cell is indicated.
(c) Model of the molecular layer and the substrate as determined from experimental STM images.
\label{intro}}
\end{figure}

Figure~\ref{intro}(a) presents an overview of a $92\times150$\,nm$^2$ area of a Ag(111) surface that is covered with a submonolayer amount of \pt.
The area exhibits a wide terrace and two substrate steps, which run nearly vertically through the image.
The steps are covered by strings of molecules, both on the upper and the lower adjacent terrace. 
However, there seems to be no clear ordering within these strings.
On the wide terrace, however, a well-ordered molecular island is observed.
The individual molecules are arranged in a hexagonal pattern and the island edges are preferentially oriented along densely packed directions of the molecular pattern.
The defect density in the island is very low (5 depressions among approximately 4700 molecules).
The large size and high degree of order of the island indicates that the regularity of the molecular arrangement leads to a significant energy gain.

Closer inspection of areas within a molecular island (Fig.~\ref{intro}(b)) reveals a hexagonal lattice of nearly circular protrusions.
Using atomically resolved images of the Ag substrate for calibration, we determined the nearest-neighbor distance to be $1.04 \pm 0.01$\,nm
and the angle between the close-packed directions of the substrate and the superlayer to be $\pm 13^\circ \pm 2^\circ$.
The centers of the molecules are located above threefold hollow sites of the Ag(111) lattice.

The experimental images suggest the model shown in Fig.~\ref{intro}(c).
This model the \pt\ molecules arranged in a corner-to-side fashion that enabled six hydrogen bonds between the TOTA platform and its six nearest neighbors.
This model is identical to the earlier proposed model for \mt\ on Ag(111).\cite{Jasper-Toennies2020}
More precisely, \mt\ forms two distinct patterns.
At low molecular densities, a honeycomb lattice was found, with triangular TOTA platforms arranged side-by-side.
This configuration enables two hydrogen bonds between adjacent molecules.
At high densities, the pattern is identical to the one observed here, with adjacent molecules arranged in a corner-to-side fashion, forming one hydrogen bond per neighbor.

\begin{figure}[h!]
\includegraphics[width=1.0\linewidth]{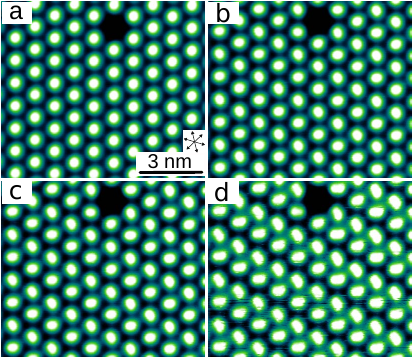}
\caption{
Topographs of a molecular island recorded at different sample voltages $V$. 
A defect (most likely a TOTA platform without phenyl ligand) near to top of the image serves as a marker.
(a) At $V=0.3$\,V, the molecules appear as nearly circular protrusions.
(b,  c) As $V$ is increased to 2.1 and 2.2\,V, the molecules develop an increasingly elliptical shape.
The long axes of the ellipses exhibit alternating orientations.
(d) At $V=2.4$\,V, the orientations of the ellipses are clearly discernible.
In addition, some indication of a constriction, possibly a nodal line, is barely visible.
Moreover, the dense molecular rows appear to have rearranged into double rows that are separated by dark groves.
\label{bias}}
\end{figure}

\subsubsection{Spectroscopic Effects on Ag(111)}

STM images of the molecular layer at negative $V$ do not show a pronounced voltage dependence, except for a gradual change in the apparent height of the island relative to the substrate.
This is consistent with the fairly featureless spectra of the differential conductance \didv\ (not shown).
However, at $V>0$ (Figure~\ref{bias}), the molecular pattern evolves in an intriguing fashion. 
While each molecule gives rise to a nearly circular protrusion at low bias (panel a), an elliptical shape becomes discernible near 1.9\,V\@.
At 2.1 and 2.2\,V (panels b and c) the ellipses are clearly visible. 
We find that the orientation of the long axes changes by $\approx 60^\circ$ between adjacent densely packed rows.
A further increase of $V$ to 2.4\,V adds new image features.
In particular, the ellipses exhibit a central constriction, which may indicate a poorly resolved nodal line.
The changes in the images within this voltage range are reflected by a peak in \didv\ (see \si), which suggests that the lowest unoccupied molecular orbital (LUMO) begins to dominate the image contrast.
As will be discussed below, the elliptical shape can be explained from the local density of states of the $\pi$-orbital that is located around the plane of the phenyl subunit. 
Therefore, the STM images reveal a unit cell with two molecules that differ by the orientation of the phenyls.

\begin{figure}[h!]
\includegraphics[width=.99\linewidth]{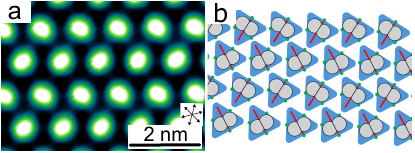}
\caption{
(a) Topograph ($V=2.2$\,V) showing the orientation of the molecular ellipses.
(b) Model of the TOTA platforms and the phenyl orientation (red lines) along with schematic $\pi$ orbitals (gray patches). 
Green dots indicate O atoms involved in bonding to the neighboring molecules.
\label{phenor}}
\end{figure}

Assuming that the image contrast above the phenyl reflects the $\pi$ orbital, the orientation of the phenyl subunits may be inferred.
Fig.~\ref{phenor} shows a detailed image of the elliptical image contrasts and a corresponding model.
It suggests that the phenyl subunits are parallel to each other along one direction of the unit cell and that they are rotated by $60^\circ$ with respect to each other along the other direction.

At slightly elevated bias voltages (Figure~\ref{bias} c and d), we find that pairs of dense molecular rows exhibit dimerization.
In this case, two closely spaced rows are separated from their neighbors by dark lines.
These lines appear $\approx50$\,pm lower that the maxima of the image at this voltage.
When the voltage is increased further, the tunneling current becomes unstable, inducing drastic changes in the topographs at 2.6\,V and above as verified by subsequent imaging at non-perturbing voltages (see \si).

\begin{figure}[h!]
\includegraphics[width=0.99\linewidth]{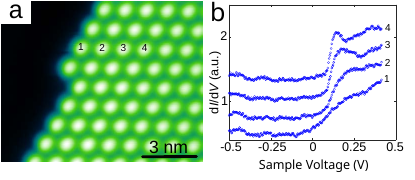}
\caption{(a) Topograph of an island edge.
The positions used for \didv\ spectroscopy are indicated by numbers.
(b) Low-bias \didv\ spectra recorded from the four molecules marked in panel a.
For clarity, the spectra have been arbitrarily shifted in the vertical direction.
The current feedback was disabled at 30\,pA and 0.5\,V\@.
\label{ssspx}}
\end{figure}

Finally, we find that the onset of the Ag(111) surface state is shifted from $-67$\,mV below the Fermi level to higher energies within the molecular islands (Fig.~\ref{ssspx}).
\footnote{We use the binding energy determined with STM, which is shifted by approximately 4\,meV compared to photoelectron spectroscopy data because of the impact of the STM tip on the local potential.\cite{Limot2003, Nicolay2001}}
Inside the island, we find a box-shaped contribution to the density of states at positive bias.
The minimum of the surface state band, which is determined from the midpoint of the rise, is shifted to $\approx 110$\,mV\@.
As expected, the rise is broadened on molecules closer to the island rim.\cite{Avouris1994, Li1999}

\subsection{Experimental Results on Au(111)}

\begin{figure}[h!]
\includegraphics[width=0.99\linewidth]{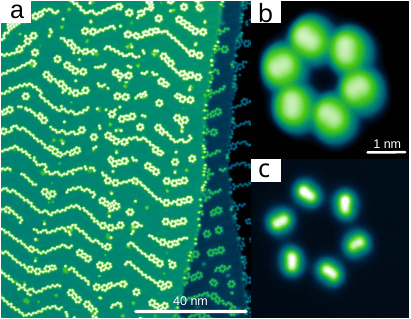}
\caption{(a) Topograph ($V=0.5$\,V) of \pt\ on Au(111).
(b) Topograph ($V=-2$,V) of an isolated hexagon of \pt\ molecules on Au(111).
(c) Constant-height image of the same hexagon ($V=-2$\,V).
\label{aupix}}
\end{figure}

We performed a brief experimental study of \pt\ on Au(111).
The herringbone reconstruction of this substrate apparently affects the pattern formation and leads to images as shown in Fig.~\ref{aupix}a.
Most of the molecules are found in strings of hexagons in the fcc areas and at elbows of the reconstruction.
At coverages approaching a monolayer, both hexagonally dense areas and honeycomb patterns were observed (not shown).
A detailed images of an isolated hexagon is shown in Fig.~\ref{aupix}b.
At the voltage used, the orientation of the phenyl subunit is readily apparent as a nodal line.
Imaging of the current at fixed tip height (Fig.~\ref{aupix}c) further confirms the data and reveals an angle of $60^\circ$ between neighboring phenyl moieties, virtually identical to the observation made on Ag(111).
We note that the phenyl pattern exhibits handedness and in fact both chiralities were observed from different hexagons.
Interestingly, the intramolecular contrast is achieved at negative voltages.
Indeed, \didv\ spectra (see \si) are featureless at positive voltages while a clear peak is observed near $-1.9$\,V\@.

\subsection{Theoretical Results}
For DFT modeling, we focused on the Ag(111) substrate because the herringbone reconstruction of the Au(111) surface adds substantial complexity. 

\begin{figure}[h!]
	\includegraphics[width=1.0\linewidth]{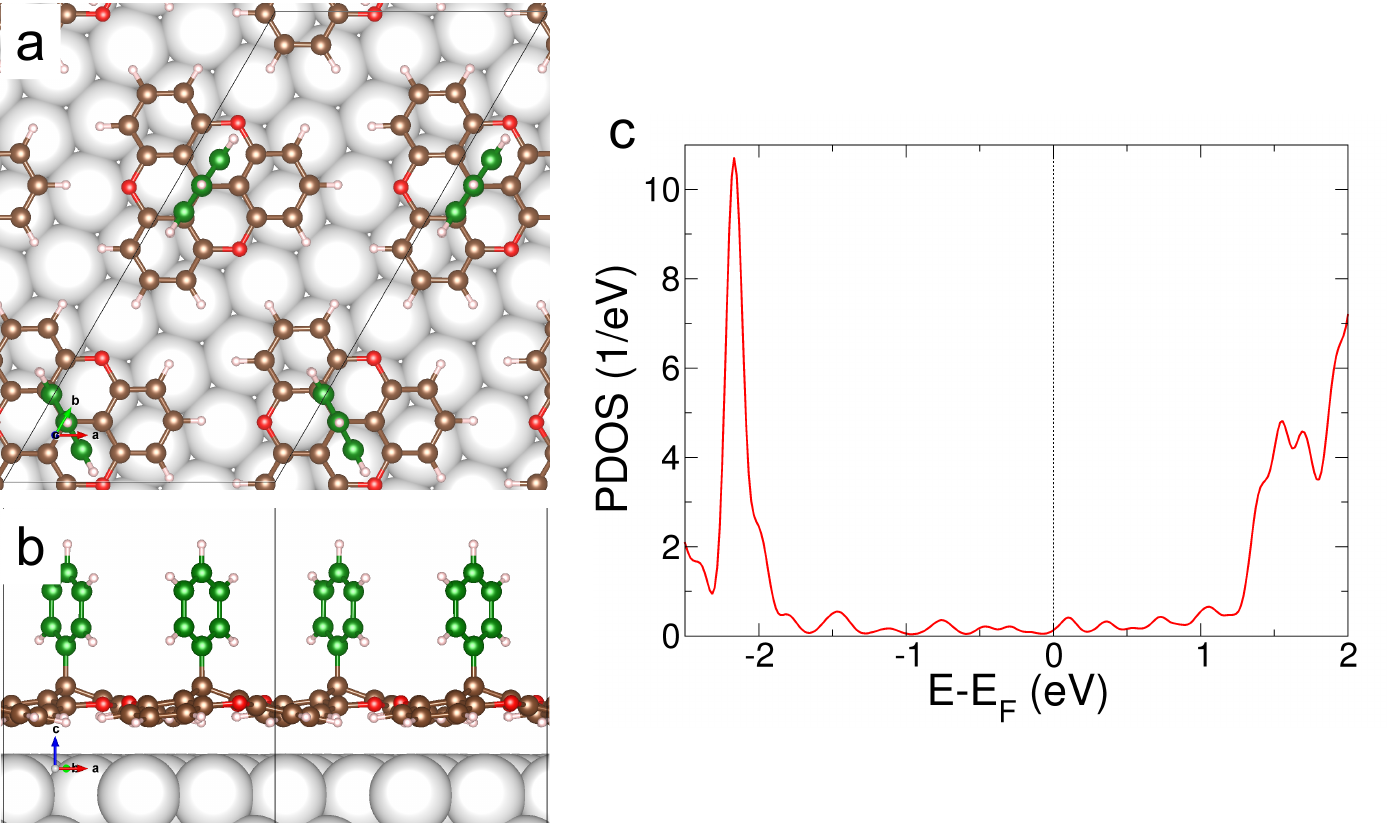}
	\caption{(a) Top and (b) side views of the most stable structure of a unit cell comprised of two \pt\ molecules on Ag(111). 
    White, brown, red, and pink spheres represent Ag, C, O, and H atoms, respectively. Green spheres represent C atoms of the phenyl moiety. 
    Black lines show the unit cell. (c) Density of states projected on the \pt\ molecule.
\label{thstruc}}
\end{figure}

Using the experimentally determined unit cell containing two inequivalent molecules we have considered different ligand orientations (see \si). The most stable configuration is shown in Figure \ref{thstruc}. As observed in the experimental data, the TOTA platforms are arranged in a hexagonal network.
The phenyl ligands are oriented along a symmetry plane of the TOTA subunit that contains an O atom and the center of the platform.
In this unit cell the phenyls are parallel to each other along the horizontal densely packed row direction.
We find that they alternate between parallel rows. 
The third symmetry equivalent orientation is not present.
These features reproduce the essential properties of the experimental structure data.

The adsorption energy of \pt\ on Ag(111) is $E_{\text{ad}}=2.238$~eV per molecule.
This energy can be decomposed in a vdW part $E_{\text{ad,vdW}}=2.584$~eV and the rest, which is repulsive by $E_{\text{ad,PBE}}=-0.346$~eV. The reason for being repulsive is the strength of the vdW interaction, which decreases the distance between the molecular layer and the surface (from ~3.4~\AA\ without including vdW interactions in the calculation to 2.8~\AA).
The molecule adopts a position unfavorable for the PBE functional and the adsorption energy goes from $E_{\text{ad,PBE}}=0.082$ to $-0.346$\,eV, which is compensated by the vdW interaction which actually binds the molecule to the surface. 

\begin{figure}[h!]
	\includegraphics[width=1.0\linewidth]{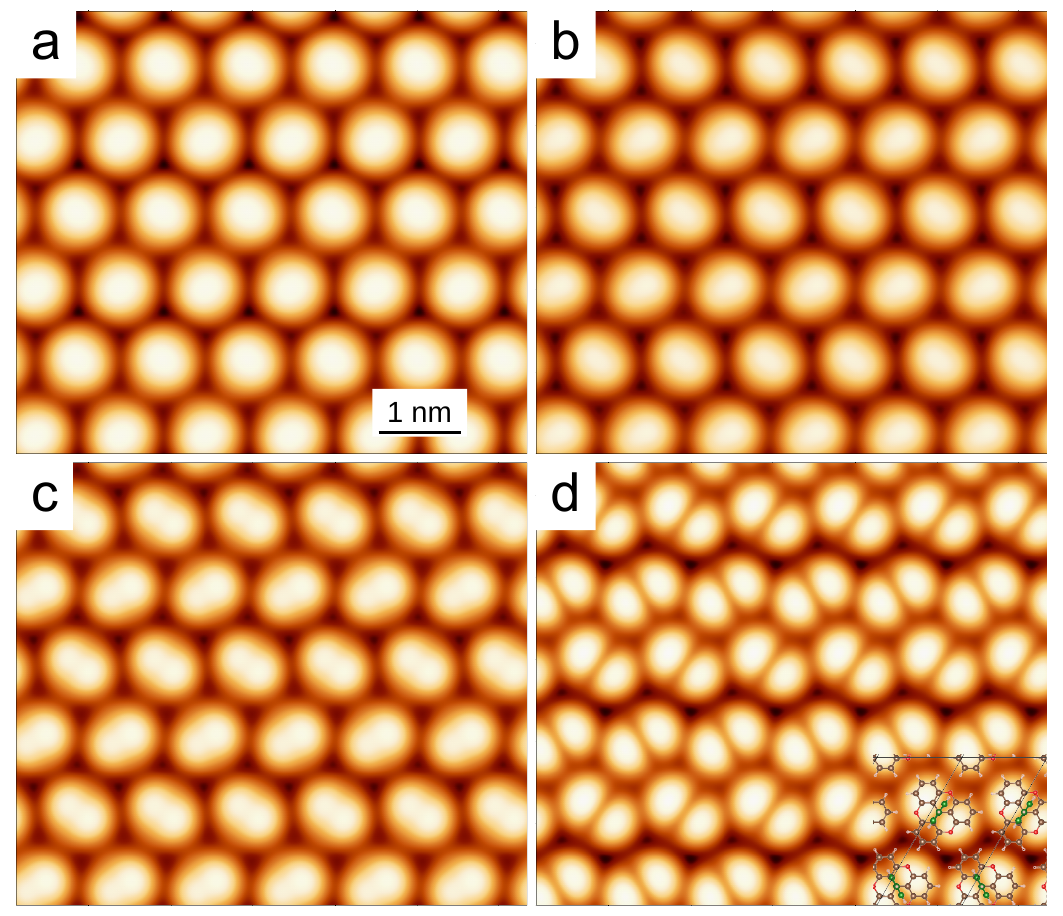}
	\caption{Calculated constant current STM images of the overlayer shown in Fig.~\ref{thstruc} for $V=$ (a) 0.5, (b) 1.1, (c) 1.2, and (d) 1.5\,V\@.
    The molecular structure is overlaid in the bottom right corner.
\label{thimg}}
\end{figure}
Next, we present constant-current images of \pt\ calculated for various voltages (Figure \ref{thimg}). 
While the molecules appear roundish at low bias, increased sample voltages lead first to an elliptical shape and finally to the appearance of a nodal plane that coincides with the phenyl plane.
This contrast at positive bias is mainly due to the lowest unoccupied orbital (LUMO), which is antisymmetric with respect to the phenyl plane. 
Other orbitals with different symmetries only contribute at larger voltages. The agreement with the experimental images at comparable voltages in Figure~\ref{bias} is quite good.
Therefore, we conclude that the orientation of the phenyl subunits can be directly determined from those experimental images that resolve the nodal plane.

\begin{figure}[h!]
	\includegraphics[width=0.8\linewidth]{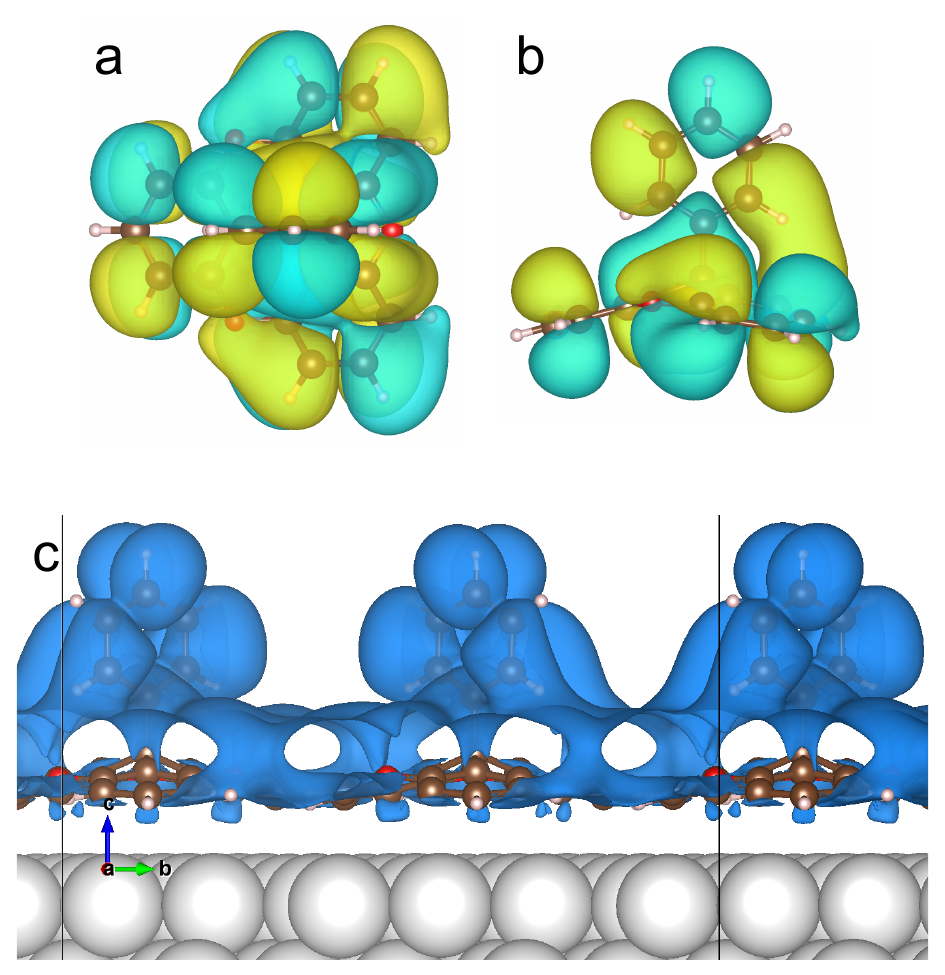}
	\caption{Isodensity contours of the LUMO of the gas-phase \pt\ viewed from (a) the top and (b) a side. (c) Partial charge density of the structure shown in Fig.~\ref{thstruc} computed between $1.37$\,V and $1.47$\,V. The charge density spills out further in the groove on the right side, appearing in the images as an apparent dimerization.
\label{thlumo}}
\end{figure}

At large voltages the molecular rows seem to dimerize: the lower areas between the horizontal phenyl rows alternate between shallow and deep grooves (brighter and darker colors). 
This effect closely resembles the experimental observations.
In order to analyze it, it is useful to inspect isodensity contours of the molecule (Figure~\ref{thlumo}).
The symmetry of the phenyl subunit is reduced by the interaction of the lower hydrogen atoms with the oxygen and the opposing phenyl subunit of the platform.
This effect is most clearly seen in the side view in Figure~\ref{thlumo}b, where the oxygen atom located on the right side of the image drastically deforms the electron cloud of the phenyl.
As a  result, the LUMO spills out further into the vacuum on the oxygen side.
Taking into account the alternation of the phenyl orientations between parallel rows, this spill out is the source of the apparent dimerization, as can be seen in Figure~\ref{thlumo}c. 
In this image, the partial charge density is shown for energies between $1.37$ and $1.47$\,V. The difference between both grooves is apparent, with the one on the right showing higher charge density, which produces a higher apparent height in the topographic images.
The different apparent heights between both grooves are visually interpreted as a dimerization.
It is worth noting that the electronic asymmetry of \pt\ has minimal effect in the geometry of the system.
For example, the tilt of the phenyl moiety with respect to the surface normal is smaller than $0.5^{\circ}$.

\subsection{Discussion}
\begin{figure}[h!]
	\includegraphics[width=1.0\linewidth]{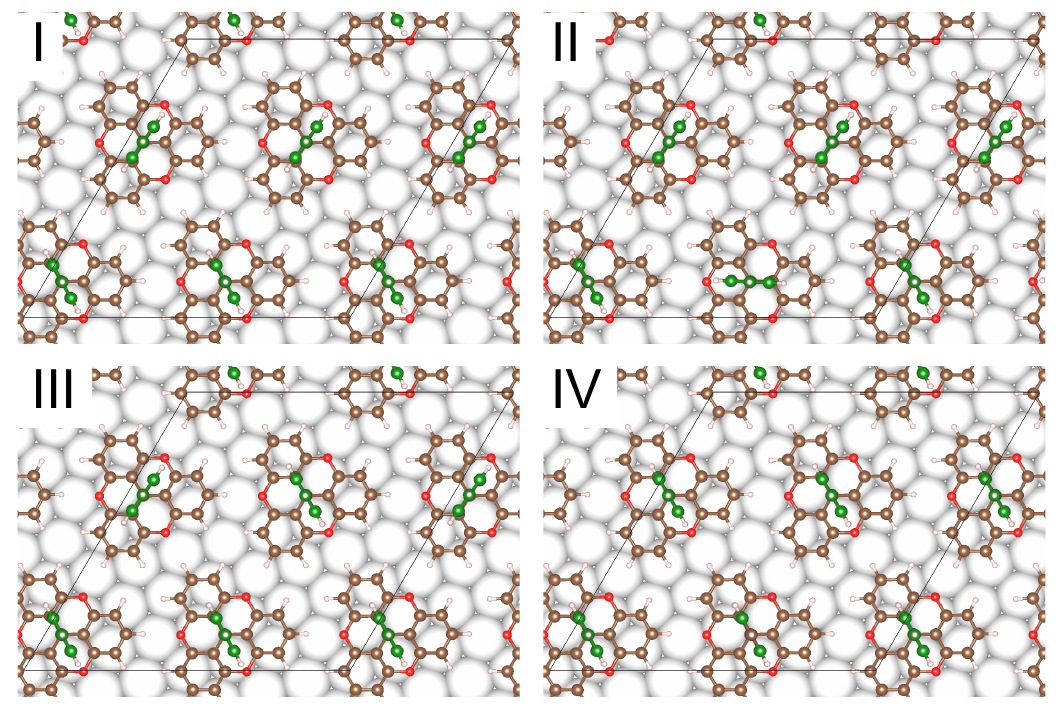}
	\caption{Top views of four configurations in a unit cell comprised of four phenyl-TOTA molecules on Ag(111).
    White, brown, red, and pink spheres represent Ag, C, O and H atoms, respectively.
    Green spheres represent C atoms of the phenyl moiety. Black lines show the unit cell.
\label{bigcell}}
\end{figure}

The orientational order of the phenyl moieties, which are separated by a distance of 1\,nm, is remarkable and deserves further discussion. 
We have shown that the orientation of the phenyl ligands can be determined experimentally and it is correctly reproduced by our model.
To shed more light on the interactions at play, we considered a larger unit cell, with doubled length in the $a$ direction.
In this model we can study configurations where phenyl moieties are not parallel in the horizontal direction.
Figure~\ref{bigcell}, presents four characteristic configurations.
The energy differences between these configurations can be found in Table~\ref{deltaE} .
The most stable configuration is I, which corresponds to the experimental observation.
Configuration II is very close in energy, actually within the estimated error margin of the calculation.
The other configurations have significantly higher energies.
Next, we consider the energy of the layer of \pt\ without the surface ($E_{\text{ph--TOTA}}$ in Table~\ref{deltaE}).
We find that the energy differences are larger, indicating that the orientational order of the phenyl ligands is not due to the interaction through the surface, which is actually detrimental.
In the \pt\ layer, only two contributions are relevant: a direct interaction between phenyl moieties, and an indirect interaction through the TOTA platforms.
To elucidate their respective roles we considered a layer of benzene molecules located at the same positions and orientations of the phenyl ligands ($E_{\text{benzene}}$ in Table~\ref{deltaE}).
$\Delta E_{\text{benzene}}$ comprises the main part of $\Delta E_{\text{ph--TOTA}}$.
Therefore, we can conclude that the principal effect is the direct interaction between phenyl moieties.
The interaction of the phenyl ligands and the TOTA platform mainly forces the ligands to adopt three possible orientations (pointing to one of the oxygen atoms in the platform).
Within this constraint, the preferred orientation is determined by the direct interaction between the phenyl moieties.
The binding energy for the benzene layer in configuration I is $E_{\text{B}}=3.84$~meV per molecule. As in the adsorption energy, the vdW part dominates with a value of $E_{\text{B, vdW}}=3.09$~meV\@.

\begin{table}
\begin{tabular}{c|ccc}
                &  $\Delta E_{\text{TOTAL}} $ & $\Delta E_{\text{ph-TOTA}}$ & $\Delta E_{\text{benzene}}$ \\
   \hline 
   I           & 0.00  &  0.00  & 0.00 \\
   II          & 0.27  &  1.85  & 1.79 \\
   III         & 4.91  &  7.68  & 6.19 \\
   IV          & 7.40  & 12.45  & 8.84 \\
\end{tabular}
\caption{Computed energy differences of the different configurations shown in \ref{bigcell} with respect to the lower energy one. $E_{\text{TOTAL}}$ is the total energy of the \pt\ layer on Ag(111); $E_{\text{ph-TOTA}}$ is the energy of the \pt\ layer without surface; and $E_{\text{benzene}}$ is the energy of a layer of benzene molecules at the same positions and orientations of the phenyl ligands. All energies are in meV.}
\label{deltaE}
\end{table}

We now go a step forward in rationalizing the interaction between the phenyl moieties.
In the hexagonal layer of benzene molecules, each molecule directly interacts with 6 neighbors.
We simplify this model by studying the interaction within a row in the unit cell of Figure~\ref{bigcell}. 
Allowing for the independent rotation of each phenyl around the surface normal we  mimic the situation of a phenyl ligand on top of the TOTA platform.
The corresponding potential energy surface (PES) as a function of the rotation angles ($\alpha, \beta$) is shown in the \si.
We can improve this model by adding the interaction of a phenyl ligand with the TOTA platform, which to first approximation is a penalty equal to 0 when the ligand is pointing to an O atom and maximum when it points in between two O atoms.
The resulting PES' is shown in the \si.
We observe minima around $0^\circ$ and $60^\circ$ and around symmetry equivalent orientations.
Looking at the ligand orientations in Figure~\ref{bigcell}, the most stable solution I maximizes the number rows with minimum energy $(0^\circ, 60^\circ)$, while only rows with higher energies are present in the least stable solution IV\@.

Our model qualitatively explains the experimental result, but is based on DFT with the PBE functional and D3 vdW corrections, a relatively simple theoretical approach.
In a benchmark of DFT methods {\it versus} state-of-the-art CCSD(T)/CBS calculations, Herman \ea\cite{Herman2023} determined that B3LYP-D3 is a good compromise between accuracy and computational demands.
We can check this theoretical approach for our system by considering the benzene-benzene interaction at long distances.
For the canonical T-shaped configuration, Czernek \ea\ \cite{Czernek2024} found an interaction energy of 4.47~meV at 0.9~nm.
Using B3LYP-D3 we obtained a value of 3.79~meV in reasonable agreement with the CCSD(T)/CBS result of Czernek \ea\@ 
We therefore computed PES and PES' using B3LYP-D3 and found the same qualitative result as for PBE-D3 (see \si).
The main difference between both theoretical methods is the interaction of the phenyl ligand with the TOTA platform, which is three times stronger with B3LYP-D3.
This will favor the tendency of the phenyl ligand to align with the O atom of the TOTA platform, but overall the interpretation of the results remains unaltered.

\section*{Conclusions}

We observed that phenyl-functionalized trioxatriangulenium (\pt) molecules self-assemble into highly ordered supramolecular networks on Ag(111) and Au(111) surfaces, exhibiting distinct orientational patterns of the phenyl moieties.
Using low-temperature STM and DFT calculations with van der Waals corrections, we resolved the orientation of individual phenyl groups and identified an intriguing dimer-like contrast effect on Ag(111), which is reproduced in the simulated STM images.
This apparent dimerization arises from an asymmetry of the phenyl wave function, induced by intramolecular hydrogen bonding between the phenyl ligand and an oxygen atom of the TOTA platform.

Our theoretical analysis reveals that the orientational order of the phenyl groups is governed primarily by direct long-range interactions between phenyl moieties, rather than substrate-mediated effects.
The TOTA platform constrains the phenyl ligands to three symmetry-equivalent orientations, and the final arrangement is determined by minimizing the intermolecular interaction energy between neighboring phenyl groups.
This interplay between intramolecular constraints and intermolecular interactions leads to the emergence of long-range orientational order and surface-induced chirality, as observed in the experiments.

These findings provide new insights into the design principles of functional molecular assemblies on surfaces, highlighting the role of subtle electronic and long-range effects in directing supramolecular organization.
The \pt\ molecule serves as a model for studying aromatic–aromatic interactions in a controlled fashion and may open an avenue for engineering surface-confined molecular architectures that enable systematic exploration of long-range molecular interactions.

\section*{Methods}

\subsection*{Experimental Details}

Experiments were carried out with a STM operated at $\approx 4.6$\,K in ultrahigh vacuum.
Ag(111) surfaces were prepared by cycles of Ar sputtering (ion energy 1.5\,keV) and annealing to 500$^\circ$C\@.
Phenyl-TOTA molecules were sublimated from a crucible heated to $\approx 100^\circ$C onto the substrate at ambient temperature.
STM tips were electrochemically etched from W wire and ion bombarded {\it in vacuo}.
STM topographs were recorded at a fixed current $I=10$\,pA unless otherwise indicated.
The bias voltage $V$ was applied to the sample.
For spectroscopy of the differential conductance \didv, a sinusoidal modulation with 10\,mV$_\mathrm{RMS}$ amplitude and a frequency between 410 and 685\,Hz was added to the bias.

\subsection*{Computational Details}

Density functional theory calculations were performed using the VASP code \cite{kresse_efficiency_1996}. The projector augmented-wave method \cite{kresse_ultrasoft_1999} was used to treat core electrons. Wavefunctions were expanded using a plane wave basis set with an energy cutoff of 500~eV. Unless noted, the PBE-D3 method was used for all calculations, consisting in the combination of the PBE exchange and correlation functional \cite{perdew_generalized_1996} complemented with the D3 method \cite{grimme_consistent_2010,grimme_effect_2011} 
to treat van der Waals interactions. Additional calculations were performed using the B3LYP-D3 method, which combines the hybrid B3LYP \cite{stephens_ab_1994} functional with the D3 method.

To simulate the experimentally determined periodicity of the molecular layer, the slab method was used with 4 layers of Ag and an in-plane unit cell determined by 
$\big(\begin{smallmatrix}
  \phantom{-}4 & 2\\
  -1 & 6
\end{smallmatrix}\big)$, 
accommodating two molecules. A $(4\times2\times1)$ $k$-grid was used to sample the Brillouin zone. The electronic structure was converged with a tolerance of $10^{-7}$, while geometries of all atoms except the two bottom Ag layers were relaxed until forces were smaller than 0.01~eV/\AA.

Adsorption energies $per$ adsorbed molecule were calculated using 
\begin{gather*}
E_{\text{ad}}[n\cdot \text{(ph-TOTA)@Ag(111)]}=[n\cdot E\text{[ph-TOTA]}+\\
E\text{[Ag(111)}]-E[n\cdot \text{ph-TOTA@Ag(111)}]/n,
\end{gather*}
where $n$ is the number of adsorbed molecules, $E\text{[ph-TOTA]}$ is the energy of one molecule, $E\text{[Ag(111)}]$ is the energy of the Ag(111) surface, and $E[n\cdot \text{ph-TOTA@Ag(111)}]$ is the energy of the full system. When computing the energies $E$ denotes the total energy, $E_{\text{vdW}}$ denotes the van der Waals part, and $E_{\text{PBE}}=E-E_{\text{vdW}}$ denotes all contributions except the vdW part. 

STM images were simulated within the Tersoff-Hamann approximation \cite{tersoff1985theory} using the method of Bocquet {\it et al.} \cite{bocquet_theory_2009} as implemented in the STMpw code \cite{lorente_stmpw_2019}.
Atomic and density plots were generated using the VESTA program \cite{momma_vesta_2011}.

\section*{Acknowledgments}

R.R.\ is grateful for financial support from projects PID2021-127917NB-I00 funded by MCIN/AEI/ 10.13039/501100011033, from project IT-1527-22 funded by the Basque Government, and from project ESiM 101046364 funded by the European Union.

\bibliography{phen}

\setcounter{figure}{0}
\renewcommand{\thefigure}{S\arabic{figure}}

\setcounter{table}{0}
\renewcommand{\thetable}{S\Roman{table}}    

\onecolumngrid

\author{Behzad Mortezapour} \affiliation{\ieap} 
\author{Sebastian Hamer} \affiliation{\ioc} 
\author{Rainer Herges} \affiliation{\ioc} 
\author{Roberto Robles} \email{roberto.robles@csic.es} \affiliation{\csic}
\author{Richard Berndt} \email{berndt@physik.uni-kiel.de} \affiliation{\ieap}

\title{\si\ for\\
Orientational Order of Phenyl Rotors on Triangular Platforms on Ag and Au(111)}

\date{\today}

\maketitle

\newpage

\section{\si: additional experimental results}
\subsection{Imaging on Ag(111) at Negative Voltages}
\begin{figure}[h!]
\includegraphics[width=.5\linewidth]{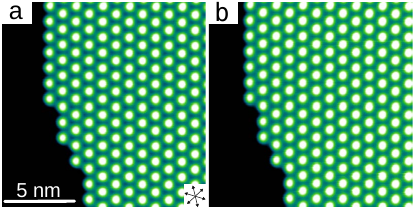}
\caption{
Topographs of a molecular island obtained at (a) $V=-0.5$ and (b) $-2.5$\,V\@.
The image contrast is hardly voltage dependent and resembles images observed at low positive voltages.
\label{negv}}
\end{figure}

\subsection{Perturbation by the Tip at Elevated Voltage}
\begin{figure}[h!]
\includegraphics[width=.5\linewidth]{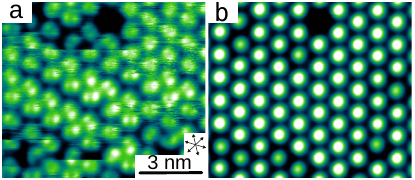}
\caption{
(a) Topograph of the surface area shown in Fig.~2 of the main text, recorded at $V=2.6$\,V\@. 
At this voltage, the tunneling current is unstable and leads to abrupt changes and streaks in the topograph. 
(b) Topograph recorded subsequently under non-perturbative conditions, $V=0.3$\,V\@. 
Some molecules appear approximately 30\,pm lower prior to the manipulation.
This effect may be due to hydrogen atoms missing from the phenyl subunit.
\label{mani}}
\end{figure}

\begin{figure}[h!]
\includegraphics[width=.5\linewidth]{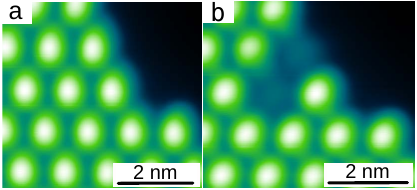}
\caption{
Removal of the phenyl ligands from two \pt\ molecules.
(a) Image of the area before manipulation.
(b) Image recorded after applying a voltage pulse to two molecules close to the image center.
During the 20\,ms pulses, the voltage was rapidly raised from 0.15 to 3.7\,V\@.
\label{phenoff}}
\end{figure}

\pagebreak
\subsection{Image of a Ag(111) Step}
\begin{figure}[h!]
\includegraphics[width=0.3\linewidth]{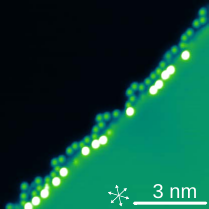}
\caption{
Topograph of a typical monatomic step of the Ag(111) substrate.
While the majority of protrusions observed at the step on both terraces are consisten with being \pt, some features appear lower, possibly due to fragments or other impurities.
\label{step}}
\end{figure}


\subsection{\didv\ Spectrum on Ag(111) at Positive Voltage}
\begin{figure}[h!]
\includegraphics[width=.35\linewidth]{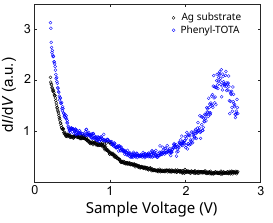}
\caption{
Spectra of the differential conductance \didv, recorded at a constant current of 100\,pA\@.
While the spectrum of the Ag(111) substrate (black line) is fairly featureless at voltages exceedind 1\,V, the spectrum of \pt\ exhibits a clear resonance centered near 2.5\,V\@.
It should be noted that features in constant-current \didv\ spectra are slightly shifted toward the Fermi level compared to constant-height data.\cite{Ziegler2009}
\label{didv}}
\end{figure}

\newpage
\subsection{Apparent Height of \pt\ on Ag(111)}

Consistent with the \didv\ spectra, there is little variation of the apparent height (Table~\ref{height}) at negative and small positive voltages and a steep increase of the height at 2\,V and above.
\begin{table}[h!]
\begin{tabular}{|l|r|r|r|r|r|r|r|r|r|r|}
\hline
$V$ (V) & $-0.5$ & $-1.0$&$-1.5$&$-2.0$&$-2.5$&	0.5&1.0  &1.5 &2.0 &2.5	\\
\hline
Height (pm) &355& 345 &340 &340 &320 &335 &360 &375 &400 &485\\
\hline
\end{tabular}
\caption{
The apparent height of \pt\ relative to the substrate recorded with $I=10$\,pA\@. 
The uncertaincies are $\pm5$\,pm except at $-0.5$\,V, where the margin is $\pm10$\,pm.
\label{height}}
\end{table}

\subsection{\didv\ Spectra on Au(111)}
\begin{figure}[h!]
\includegraphics[width=0.7\linewidth]{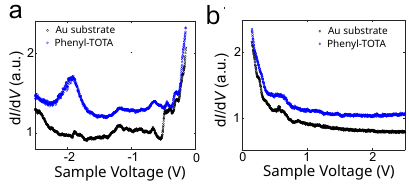}
\caption{Constant-current ($I=50$\,pA) \didv\ spectra of \pt\ on Au(111) (blue lines) and of the substrate (black lines).
(a) Negative voltages.
(b) Positive voltages.
\label{auspx}}
\end{figure}

\pagebreak
\section{\si: additional computational results}
\subsection{Determination of the Adsorption Sites of an Isolated \pt\ molecule}\begin{figure}[h!]
\includegraphics[width=0.7\linewidth]{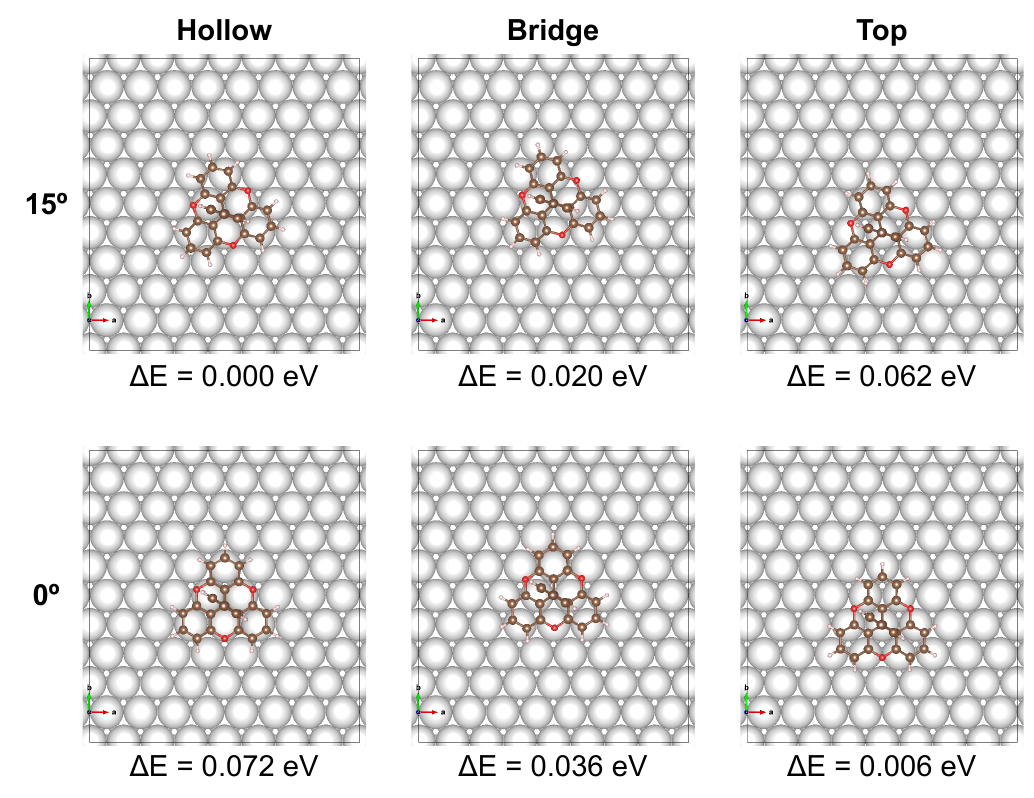}
\caption{Six geometries of an isolated \pt\ molecule on Ag(111).
$0^\circ$ and $15^\circ$ rotations with respect to the $(1\bar{1}0)$ direction are considered.
The center of the molecule is placed at hollow, bridge or top positions.
Gray, brown, red, and pink spheres represent Ag, C, O and H atoms, respectively. Black lines show the unit cell.
The energy differences of all configurations are indicated below each panel.
\label{positions}}
\end{figure}
To determine the most stable configuration of \pt\ molecules on Ag(111) we considered a rectangular (8$\times$5) unit cell with just one isolated molecule. 
The experiments indicate a $\approx15^\circ$ rotation of the molecules with respect to the $(1\bar{1}0)$ densely packed direction on the (111) surface.
We then considered $0^\circ$ and $15^\circ$ rotations of the molecules and hollow, bridge and top positions for the center of the molecule (Figure~\ref{positions}).
The most stable configuration corresponds to a molecule rotated by $15^\circ$ at the hollow position.
We used this geometry for the construction of the unit cells of the monolayer.

\newpage

\subsection{Simulation of Different Ligand Orientations}
\begin{figure}[h!]
\includegraphics[width=0.9\linewidth]{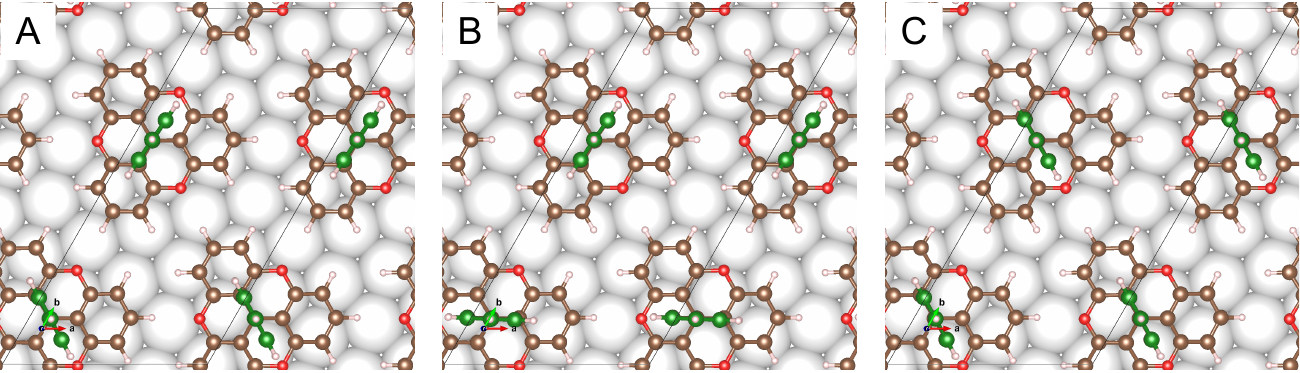}
\caption{Three different orientations of the phenyl ligands in the cell containing two inequivalent \pt\ molecules.
Configuration A is discussed in the main text. White, brown, red, and pink spheres represent Ag, C, O and H atoms, respectively.
Green spheres represent C atoms of the phenyl moiety. Black lines show the unit cell.
\label{fig_cell}}
\end{figure}

\begin{figure}[h!]
\includegraphics[width=0.75\linewidth]{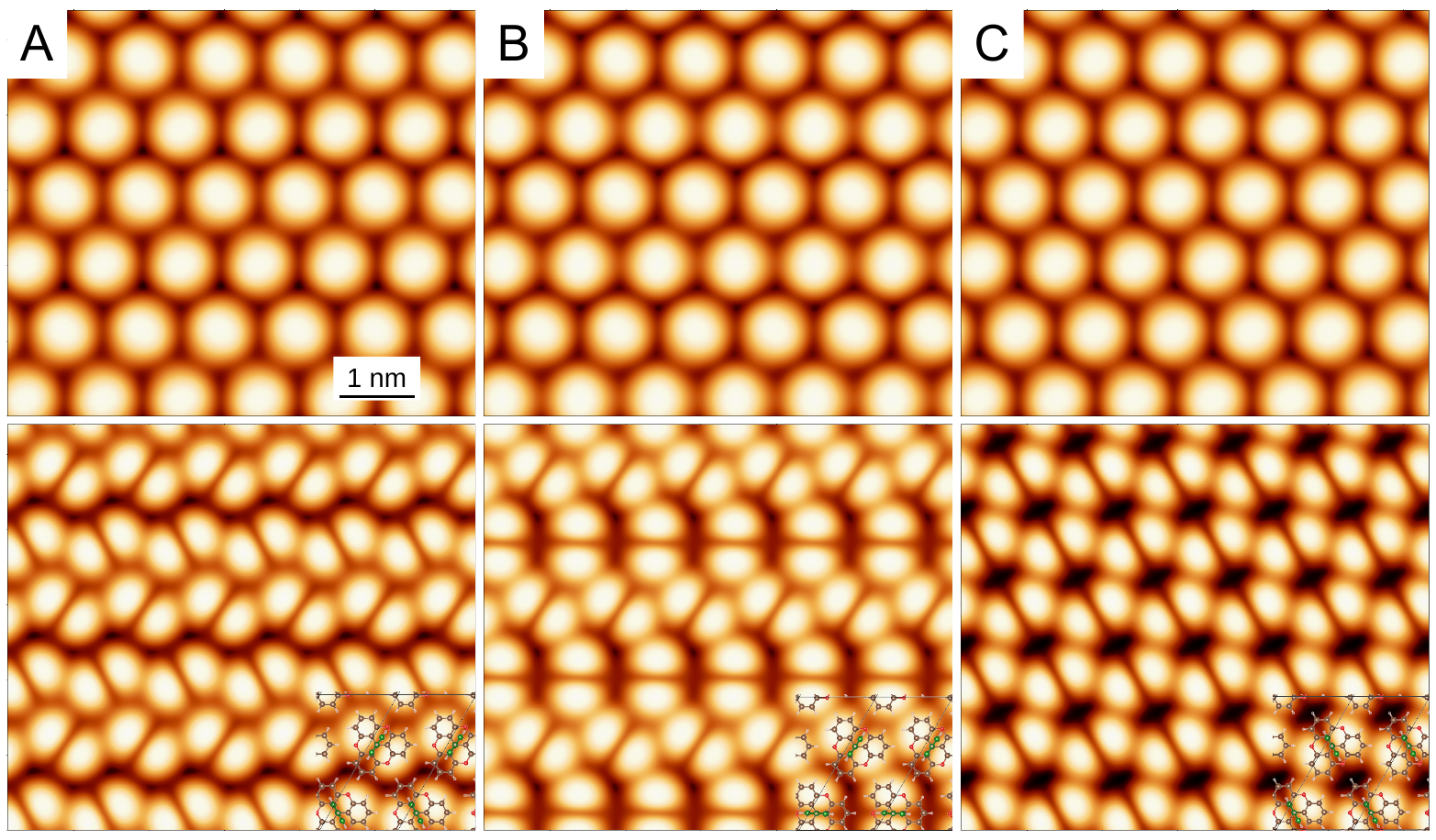}
\caption{Constant-current STM images simulated for the three configurations of Figure~\ref{fig_cell}.
The first row was computed for $V=0.5$\,V\@.
The images are essentially featureless and independent of the phenyl orientation.
The second row at $V=1.6$~V captures the LUMO and reveals the different orientations of the phenyl moieties.
The molecular structure is overlaid in the bottom images.
\label{fig_STM}}
\end{figure}

We considered different orientations of the phenyl ligands, which are shown in Figure~\ref{fig_cell}.
The corresponding energy differences and adsorption energies are shown in Table~\ref{energies}.
The energy differences between the configurations are small, of the order of meV\@.
However, as seen in Figure~\ref{fig_STM}, the simulated STM images at $V=1.6$~V are quite different and only configuration A is compatible with the experimental observations. 

The properties of this system are mainly governed by vdW interactions.
Many approaches to include them are suggested in the literature.
We have tested five of them: two are semiempirical pairwise corrections (PBE-D3 \cite{grimme_consistent_2010,grimme_effect_2011} and PBE+vdW$^{\text{surf}}$ \cite{ruiz_density_2016}), while the other three are nonlocal vdW-DF functionals in the spirit of Dion {\it et al} \cite{dion_van_2004}.
Regardless of the approach, configuration A is found to be the ground state, while the other two configurations are within few meV\@.
The variation of the adsorption energies is wider, going from 1.6 to 2.2~eV per molecule.
The van der Waals part of the adsorption energy dominates in all cases.
The distance between the molecule and the surface is similar for all methods, varying from 2.7 to 2.9~\AA\@.
For reference, the distance for the PBE functional, without including any vdW correction, is of 3.4~\AA\@.
In summary, the same qualitative picture emerges from all methods and the determination of the experimental configuration is consistent for all of them.

\begin{table}[h!]
\setlength{\tabcolsep}{8pt}
\begin{tabular}{l|ccc|ccc|ccc|c}
\hline \hline
 & \multicolumn{3}{c|}{\bf{A}} & \multicolumn{3}{c|}{\bf{B}} & \multicolumn{3}{c|}{\bf{C}} &  \\
& $\Delta E$ & $E_{\text{ad}}$ &  $E_{\text{ad,vdW}}$ & $\Delta E$ & $E_{\text{ad}}$ & $E_{\text{ad,vdW}}$ & $\Delta E$ & $E_{\text{ad}}$ &  $E_{\text{ad,vdW}}$ & $\;d\;$ \\
\hline
PBE-D3 \cite{grimme_consistent_2010,grimme_effect_2011} & 0.00 & 2.238 & 2.584 & 2.47 & 2.237 & 2.588 & 3.69 & 2.236 & 2.594 & 2.8 \\
PBE+vdW$^{\text{surf}}$\cite{ruiz_density_2016} & 0.00 & 2.140 & 2.735 & 0.40 & 2.139 & 2.760 & 3.72 & 2.138 & 2.753 & 2.7 \\
Hamada \cite{hamada_van_2014} & 0.00 & 1.649 & 2.886 & 3.33 & 1.647 & 2.892 & 3.18 & 1.647 & 2.893 & 2.9 \\
CX \cite{berland_exchange_2014} & 0.00 & 1.954 & 1.759 & 3.65 & 1.952 & 1.752 & 3.93 & 1.952 & 1.771 & 2.8 \\
optB86b \cite{klimes_van_2011} & 0.00 & 2.114 & 3.777 & 2.97 & 2.113 & 3.783 & 3.11 & 2.113 & 3.792 & 2.9 \\

\hline \hline
\end{tabular}
\caption{
Energy differences $\Delta E$ (in meV), adsorption energies $E_{\text{ad}}$ and $E_{\text{ad,vdW}}$ (in eV per molecule) and distances $d$ (in \AA) between the molecule and the Ag(111) surface. They are shown for the three configurations in figure~\ref{fig_cell} and for five different theoretical approaches.
\label{energies}}
\end{table}

\pagebreak

\subsection{Chain of Benzene Molecules}

\begin{figure}[h!]
\includegraphics[width=0.80\linewidth]{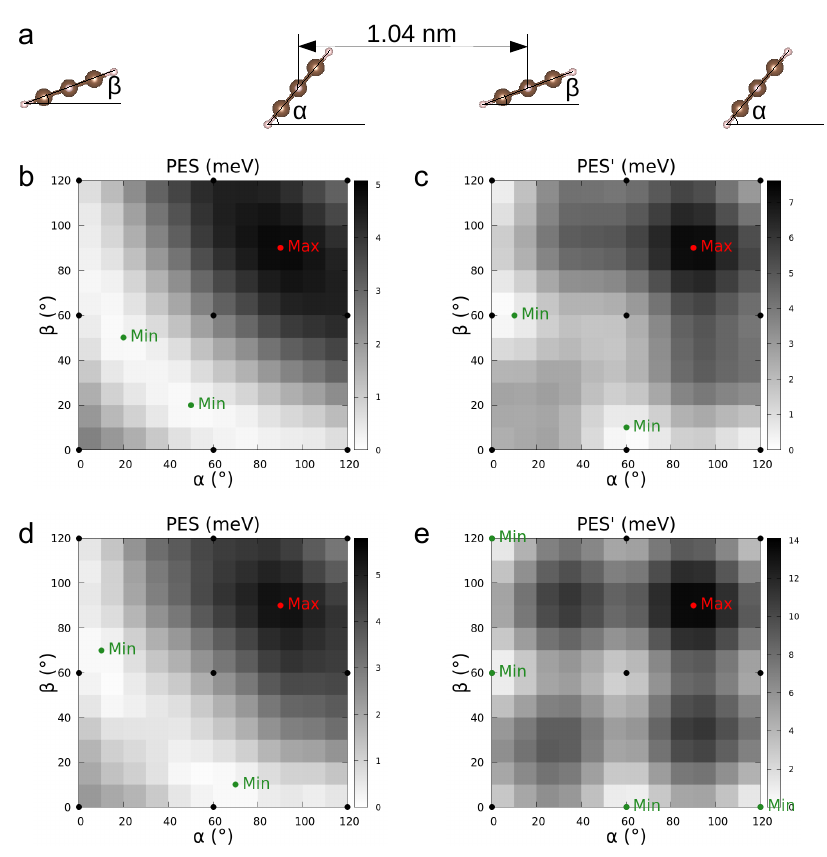}
\caption{
(a) Schematic top view of a chain of benzene molecules with two inequivalent molecules rotated by angles $\alpha$ and $\beta$, respectively.
The molecular centers are separated by 1.04~nm.
The model simulates the direct long-range interaction of the phenyl ligands on the TOTA platform.
(b) PBE-D3 potential energy surface (PES) of the model in (a) as a function of the angles ($\alpha$, $\beta$).
Minimum (maximum) energy points are marked in green (red).
Black dots indicate multiples of $60^\circ$.
(c) PBE-D3 PES' including the interaction of a phenyl ligand in the TOTA platform for the given angle.
(d--e) PES as (b) and (c), but calculated with the B3LYP-D3 functional and vdW correction.
\label{fig_benzenes}}
\end{figure}

We studied the direct interaction between phenyl moieties using the model shown in Figure~\ref{fig_benzenes}a.
It consists of a chain of benzene molecules at the same distance as the phenyl moieties in the cell of Fig.~\ref{fig_cell}.
The benzenes are constrained to rotate in a vertical axis mimicking the phenyl ligands of \pt.
We considered two inequivalent molecules rotated by angles $\alpha$ and $\beta$. We varied these angles from $0^\circ$ to $120^\circ$ in $10^\circ$ intervals and computed the energy differences to obtain the PES as a function of ($\alpha$, $\beta$) (Figure~\ref{fig_benzenes}b).
The minimum energy is found at ($20^\circ, 50^\circ$) and corresponds to a structure resembling the global minimum of a benzene dimer, namely a tilted T-shaped configuration \cite{Czernek2024}.
There is ample region with an energy within 1~meV of the minimum.
The maximum energy configuration is 5.04~meV higher in energy and corresponds to $(90^\circ, 90^\circ)$, with the benzenes parallel and facing each other.

The model can be improved by adding the energy of rotating a phenyl ligand on top of the TOTA platform.
This energy is minimal for multiples of $60^\circ$, when the phenyl ligand is pointing to an oxygen atom, and maximum for $30^\circ$, when the ligand is parallel to one side of the TOTA platform.
We add to the PES the values for each angle going from 0~meV for the minimum to 1.39~meV for the maximum, obtaining the modified PES' in Figure~\ref{fig_benzenes}c.
The minima are around $(0^\circ, 60^\circ)$ and symmetry equivalent combinations $(60^\circ, 0^\circ)$, $(0^\circ, 120^\circ)$ and $(120^\circ, 0^\circ)$.
Using the results in PES' we can rationalize the most stable configuration in the layer of \pt\ molecules shown in Figure~\ref{fig_cell}A as the one that maximizes the number of rows with the minimum energy configuration.
Conversely, for configuration C none of the rows corresponds to a orientation of phenyl ligands similar to the minimum energy one.

The same analysis was done for the B3LYP-D3 functional, which according to Herman \ea\ \cite{Herman2023} is good choice for studying benzene--benzene interactions.
The results are shown in Figure~\ref{fig_benzenes}d and e.
For the PES, a similar shape is obtained.
Now the maximum at $(90^\circ, 90^\circ)$ is 5.79~meV higher than the minimum at $(10^\circ, 70^\circ)$.
This value is not far from the 5.04~meV value found for PBE-D3.
The fact that the PES's computed with these two methods are similar reflects a comparable long-range interaction due to the van der Waals forces.
However, the interaction with the TOTA platform is stronger within B3LYP-D3 (between 0 and 4.36~meV) compared to 1.39~meV for PBE-D3.
As can be seen in PES' (Figure~\ref{fig_benzenes}e), this stronger interaction further stabilizes the configurations at $(0^\circ, 60^\circ)$ and in general configurations where $\alpha$ and $\beta$ are multiples of $60^\circ$.
This fact reinforces the agreement with the experimental observations, where the phenyl ligands are only observed with angles multiple of $60^\circ$.
In conclusion, the intramolecular interaction of the phenyl moieties on the TOTA platform is the major difference between the PBE-D3 and the B3LYP-D3 methods.
Overall, the same qualitative interpretation of the experimental results is achieved with both theoretical methods.

\bibliography{phen}

\end{document}